\documentstyle[fleqn,12pt]{article}

\begin{document}
october 1996 \hfill preprint cond-mat

\vspace{2cm}
\Large
\centerline{Spectral Functions for the Holstein Model}
\vspace{1.5cm}
\centerline{J. M. Robin}
\vspace{0.5cm}
\normalsize
\centerline{Centre de Recherche sur les Tr\`es Basses Temp\'eratures,}
\centerline{25, avenue des Martyrs, BP 166, 38042 Grenoble Cedex 9, France.}
\vspace{1cm}
\centerline{\today}
\vspace{1cm}

\baselineskip 22pt

\begin{abstract}
We perform an unitary transformation for the
symmetric phonon mode of the Holstein molecular crystal hamiltonian.
We show how to compute the electronic spectral functions by exact
numerical diagonalisation of an effective hamiltonian fully taking
account of the symmetric phonon mode, usually discarded.
\end{abstract}

\vspace{1.5cm}

This paper explains carefully how to simplify numerical investigations
of the Holstein model, and other related model containing electron-phonon
interactions, using exact diagonalization.
Such a simplification was introduced many years ago by Ranninger and
Thibblin for the two sites polaron problem\cite{Rann92}.
The symmetric phonon mode is picked out and taken account analytically
as displaced oscillator, reducing the number of phonon modes by one
(one is a large number for small clusters).
Since then, this trick has been widely used. However, many
authors\cite{Alex94},
neglected important
contributions from the symmetric mode.

We consider the Holstein hamiltonien, without Coulomb repulsion, given
by
\begin{equation}
H \; = \; \varepsilon_{0} \sum_{j,\sigma} c_{j,\sigma}^{\dagger} c_{j,\sigma}
\; - t \; \sum_{j,\delta,\sigma} c_{j+\delta,\sigma}^{\dagger} c_{j,\sigma}
\; + \; \omega_{0} \sum_{j} a_{j}^{\dagger} a_{j} \;
- \; g \omega_{0} \; \sum_{j,\sigma} c_{j,\sigma}^{\dagger} c_{j,\sigma}
( a_{j} + a_{j}^{\dagger} ).
\end{equation}
The sum over $\delta$ is on the $z=2d$ nearest neighbours ($z=1$ for a two
sites lattice) and $\varepsilon_{0}$ is some parameter.
For a $M$ sites lattice, the symmetric phonon mode  is given by
$a_{s} = (a_{1} + a_{2} + \ldots + a_{M})/\sqrt{M}$ and the corresponding
hamiltonian is given by
\begin{equation}
H_{s} \; = \; \omega_{0} a_{s}^{\dagger} a_{s} \; - \;
\frac{g \omega_{0} N}{\sqrt{M}} \; (a_{s}^{\dagger} + a_{s}),
\end{equation}
where $N$ is the total number of electrons. The hamiltonian becomes
$H = H_{e} + H_{s}$. $H_{e}$ is an effective hamiltonian with one phonon
mode missing which has to be numerically solved. Indroducing new
bosonic operators $b_{N} = a_{s} - gN/\sqrt{M}$, we obtain the diagonal
form
\begin{equation}
H_{s} \; = \; \omega_{0} b_{N}^{\dagger} b_{N} \; - \;
g^{2} N^{2} \omega_{0} / M .
\end{equation}
Let
$| \; n, N )$ be the eigenstates of $H_{s}$ with eigenvalues
$E_{n}^{s, N} = n\omega_{0} - g^{2}N^{2}\omega_{0}/M$.
This transformation is an unitary transformation,
$b_{N} \; = \; U_{N}\;a_{s}$ with
\begin{equation}
U_{N} \; = \; e^{g N ( a_{s}^{\dagger} - a_{s} ) / \sqrt{M}}.
\end{equation}
The coherent states are given by
$| \; n, N ) \; = \; U_{N}\;|\;n>$ where $|\;n>$  are the eigenstates of
$a_{s}^{\dagger}a_{s}$.
The matrix elements are given by
$(n',N'\;|\;n,N) \; = \; <n'\;|\;U_{N-N'}\;|\;n>$.
In the particular case of a transition from the ground state with $N$
electrons to an excited state with $N'$ electrons, we get
\begin{equation}
( n', N' \; | \; 0, N ) \; = \frac{1}{\sqrt{n'!}}
\; e^{-\frac{g^{2} (N-N')^{2} }{2 M} } \;
\left(\frac{g^{2} (N-N')^{2} }{M} \right)^{n'/2}.
\end{equation}
We now show how this matrix element arise in the expression of the
electronic correlation functions, say
$J_{1}(t) = \; < c_{k,\sigma}(t) \; c_{k,\sigma}^{\dagger} >$
where $k$ is the momentum.
Let $|\;m,N>$ be the eigenstates of $H_{e}$ with eigenvalues
$E_{m}^{e, N}$. At $T=0$, we get
\[
J_{1}(t) \; = \; <0,N\;|\; (0,N\;|\;c_{k,\sigma}(t) \;
c_{k,\sigma}^{+}\;|\;0,N>  \;|\;0,N)
\]
\[
\; \; \; \; \; \; \; \; \; \; = \;
\sum_{m,n}  <0,N\;|\; (0,N\;|\;c_{k,\sigma}(t)\;
|\;m,N+1>  \;|\;n,N+1) \; \times
\]
\begin{equation}
\; \; \; \; \; \; \; \; \; \; \; \; \; \; \; \;  \times \;
<m,N+1\;|\; (n,N+1\;|\;c_{k,\sigma}^{\dagger}\;|\;0,N> \;|\;0,N)
\end{equation}
and the spectral function is given by
\[
J_{1}(\omega) \; = \; 2 \pi \sum_{n,m} \; \left|\; (n,N+1\;|\;0,N)\;
\right|^{2} \; \left|\;  < 0,N\;|\; c_{k,\sigma} \;|\;m,N+1 >
\right|^{2} \; \times
\]
\begin{equation}
\; \; \; \; \; \; \; \; \; \; \; \; \; \; \; \; \; \; \; \; \; \times \;
\delta ( \omega + E_{0}^{s,N} + E_{0}^{e,N} - E_{n}^{s,N+1} - E_{m}^{e,N+1} )
\label{J1true}
\end{equation}
This result is to be compared with the na\"{\i}ve expression based only
on $H_{e}$ which is
\begin{equation}
J_{1}^{\mbox{na\"{\i}ve}}(\omega) \; = \; 2 \pi \sum_{m}
\left|\;  < 0,N\;|\; c_{k,\sigma} \;|\;m,N+1 >
\right|^{2} \; \delta ( \omega + E_{0}^{e,N} - E_{m}^{e,N+1} ).
\label{J1false}
\end{equation}
Figure 1 shows $J_{1}(k=0,\omega)$ with $\omega_{0}/t=0.2$ and
$\lambda=g^{2}\omega_{0}/t=1.2$ for the two sites problem with ground
state corresponding to $N=0$.
Part (a) reproduces Alexandrov et als result
(figure 4.a of reference\cite{Alex94}) using equation (\ref{J1false}),
while part (b) shows the correct result using equation (\ref{J1true}).
We used $50$ phonon states for $H_{e}$ and $20$ phonon states for $H_{s}$.
The exact result for $g=0$ is\cite{Mahan}
\begin{equation}
J_{1}^{g=0}(k=0, \omega) \; = \; 2 \pi \sum_{\ell = 0}^{\infty} \;
e^{g^{-2}} \; \frac{g^{2 \ell}}{\ell!} \;
\delta ( \omega - \varepsilon_{0} - E_{_P} - \ell \omega_{0} ),
\end{equation}
with $E_{_P}=g^{2}\omega_{0}$, an useful test formula.

\vspace{2cm}

I thank J. Ranninger for interesting discussions.

\vspace{2cm}
\Large
\noindent
{\bf Figure Captions}
\normalsize
\vspace{1cm}

{\bf Figure 1}: Spectral function $J_{1}(k=0,\omega)$ for the two sites Holstein
Hamiltonian  with $\omega_{0}/t=0.2$ and $\lambda=1.2$.
(a) shows the na\"{\i}ve result obtained without the symmetric contribution
while (b) shows the correct result.

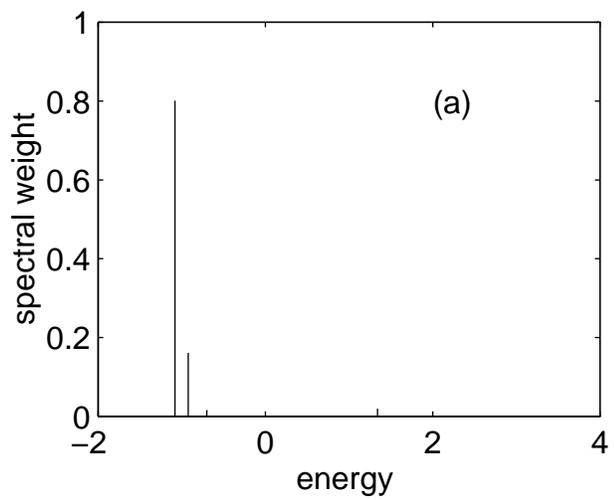
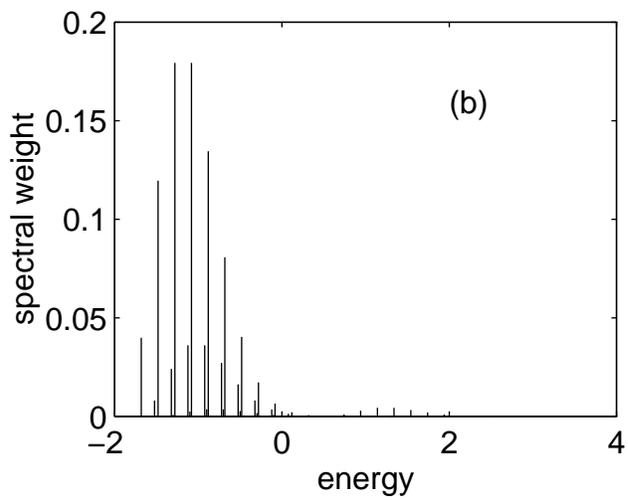
\end{document}